\documentclass[11p]{article}

\usepackage[utf8]{inputenc}
\usepackage{amsmath,amssymb}
\usepackage{mathtools}
\usepackage{epsfig,graphicx}
\usepackage{siunitx}
\usepackage{xcolor}
\usepackage[colorlinks]{hyperref}
\usepackage{lipsum}
\usepackage{multirow}
\usepackage{authblk}
\usepackage{cite}
\usepackage[left=2.5cm,right=2.5cm,top=2.5cm,bottom=2.5cm]{geometry}

\newcommand{\bra}[1]{\langle #1 |}
\newcommand{\ket}[1]{| #1 \rangle}
\newcommand{\braket}[2]{\langle #1 | #2 \rangle}

\newcommand{\HN}{Rytova-Keldysh }

\renewcommand{\vec}[1]{\boldsymbol{#1}}

\begin{document}


\title{Single-particle properties of topological Wannier excitons in bismuth chalcogenide nanosheets} 


\author[1]{L. Maisel Licer\'an \textsuperscript{$*$}}
\author[1]{F. Garc\'ia Fl\'orez \textsuperscript{$\dagger$}}
\author[2]{L. D. A. Siebbeles \textsuperscript{$\ddag$}}
\author[1]{H. T. C. Stoof \textsuperscript{\S}}

\affil[1]{Institute for Theoretical Physics and Center for Extreme Matter and Emergent Phenomena, Utrecht University, Princetonplein 5, 3584 CC Utrecht, The Netherlands}
\affil[2]{Optoelectronic Materials Section, Department of Chemical Engineering, Delft University of Technology, Van der Maasweg 9, 2629 HZ Delft, The Netherlands}

\date{}
\maketitle

\vspace{-1cm}

\begin{center}
    \textsuperscript{$*$}\href{mailto:l.maiselliceran@uu.nl}{l.maiselliceran@uu.nl}\\
    \textsuperscript{$\dagger$}\href{mailto:f.garciaflorez@pm.me}{f.garciaflorez@pm.me}\\
    \textsuperscript{$\ddag$}\href{mailto:l.d.a.siebbeles@tudelft.nl}{l.d.a.siebbeles@tudelft.nl}\\
    \textsuperscript{\S}\href{mailto:h.t.c.stoof@uu.nl}{h.t.c.stoof@uu.nl}
\end{center}

\vspace{0.25cm}

\begin{abstract}
	We analyze the topology, dispersion, and optical selection rules of bulk Wannier excitons in nanosheets of Bi\textsubscript{2}Se\textsubscript{3}, a topological insulator in the family of the bismuth chalcogenides.
    Our main finding is that excitons also inherit the topology of the electronic bands, quantified by the skyrmion winding numbers of the constituent electron and hole pseudospins as a function of the total exciton momentum.
    The excitonic bands are found to be strongly indirect due to the band inversion of the underlying single-particle model.
    At zero total momentum, we predict that the $s$-wave and $d$-wave states of two exciton families are selectively bright under left- or right-circularly polarized light.
    We furthermore show that every $s$-wave exciton state consists of a quartet with a degenerate and quadratically dispersing nonchiral doublet, and a chiral doublet with one linearly dispersing mode as in transition metal dichalcogenides.
    Finally, we demonstrate the existence of topological edge states of chiral excitons arising from the bulk-boundary correspondence.
\end{abstract}


\section{Introduction}

Three-dimensional topological insulators, and all other topological materials for that matter, are presently receiving much attention because of their excellent prospects for energy-efficient electronics, (pseudo)spintronics devices, and quantum information processing \cite{fu2007,hasan2010,qi2011,moore2010,yan2012,ando2013,ando2015,shen2012,ortmann2015,asboth2016short,tkachov2015topological,luo2019,cha2013,hu2020,kou2017,yue2018,liu2019,hohenadler2013}.
Prototypical examples of three-dimensional topological insulators are the bismuth chalcogenides Bi$_2$Se$_3$ and Bi$_2$Te$_3$.
Since in linear response these materials are ideally conducting only due to the presence of massless Dirac fermions on their surface, most experiments with topological insulators have focused on these unusual topologically protected surface states.
However, the situation changes dramatically upon photoexcitation, as excitons and unbound charges may be produced in the bulk with a topologically nontrivial band structure.
Consequently, it is important for the understanding of light-matter interactions to investigate also the bulk properties of topological insulators and in particular the precise topological nature of the excitons, whose presence or absence is crucial in optoelectronic devices such as lasers \cite{ye2015a,paik2019,schneider2013,wen2020}, light-emitting diodes \cite{tsintzos2008,bajoni2008,uoyama2012}, and photovoltaic cells \cite{adachi2001,gregg2003,luther2019,congreve2013}.
In the context of quantum information processing a particularly interesting question is if the exciton topology is transferred to the quantum state of the photons emitted via photo- or electro-luminescence \cite{kung2019observation}.
Apart from such applications, the many-body physics of topological excitons is thought to be very exciting and accessible experimentally by well-established pump-probe techniques.
Important examples of interesting many-body states are the topological excitonic insulator \cite{du2017,khatibi2020a,pereira2021topological,jia2021evidence,sun2021evidence} and a Bose-Einstein condensate of topological excitons or possibly even biexcitons \cite{garciaflorez2020a}.

In recent years, therefore, there has been an increasing interest in the study of excitons formed in topological insulators.
Semiclassically and within the effective-mass approximation, a pioneering approach has been to introduce Berry-phase corrections to the electron-hole interaction due to the topological band structure and determine the exciton energy spectrum \cite{hichri2019,trushin2018,zhou2015berry,srivastava2015a,sablikov2017,allocca2018}.
The approach presented in these works mostly focuses on the case of a total exciton momentum $\vec{Q} = \vec{0}$, which is sufficient for studying optically active excitons, but not enough to obtain their global topological properties.
The latter has been achieved for Frenkel excitons within a Hubbard-like model with on-site Coulomb interactions\cite{chen2017a}. However, because of the insulating nature of the bulk of bismuth chalcogenide nanosheets and their geometry, which leads to a lower-than-bulk dielectric constant due to the surrounding medium, we expect in these materials long-ranged electron-hole interactions allowing for the formation of so-called topological Wannier excitons \cite{wannier1937,chernikov2014}.
In this article we therefore study the concrete example of (quasi-)two-dimensional bulk excitons in Bi\textsubscript{2}Se\textsubscript{3} nanosheets.

\section{Results}

\subsection{Band structure of Bi\textsubscript{2}Se\textsubscript{3} nanosheets}

\begin{figure}[t]
    \begin{center}
        \includegraphics[width=0.6\linewidth]{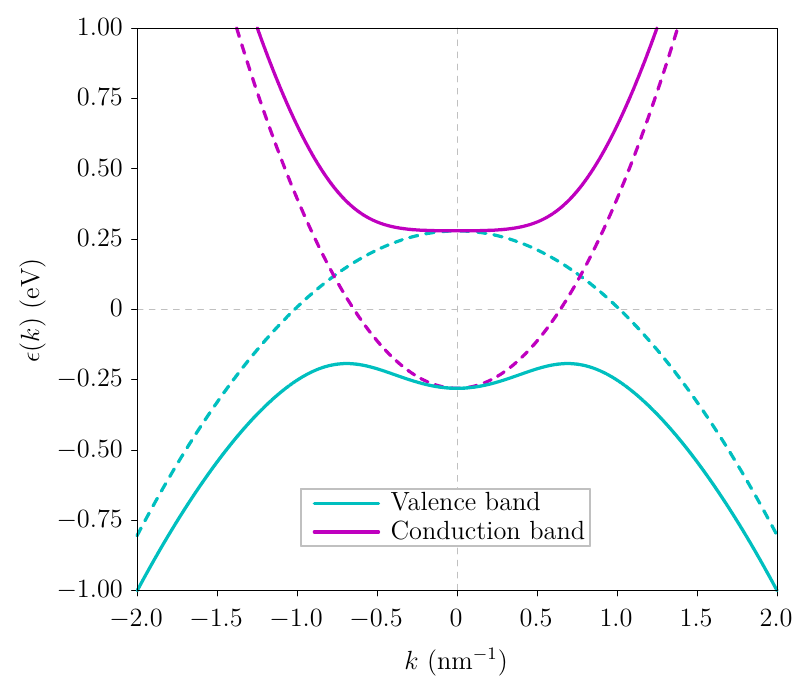}
        \caption{
            \textbf{Band structure of two-dimensional Bi$_2$Se$_3$ around the $\vec{\Gamma}$ point.}
            Both the conduction bands and the valence bands are shown in solid lines.
            Each is two-fold degenerate, and due to spin-orbit coupling features a band inversion and an avoided crossing of the corresponding uncoupled states depicted with dashed lines.
            The dispersions are isotropic in the $\vec{k} \boldsymbol{\cdot} \vec{p}$ approximation we are using.
        }
        \label{fig:eh_bands}
    \end{center}
\end{figure}

Because we are only interested in the physics around the $\Gamma$ point, we restrict ourselves to the bands closest to the Fermi surface.
For all numerical purposes we choose a nanosheet thickness of $\SI{6}{\nano\meter}$, \textcolor{black}{which approximately corresponds to 6 quintuple layers (QLs)}.
This is sufficiently large to preserve the nontrivial bulk topology, but also sufficiently small for the problem to be mainly regarded as two-dimensional.
The effective Hamiltonian governing the physics around the $\Gamma$ point in Bi\textsubscript{2}Se\textsubscript{3} nanosheets is equivalent to the BHZ Hamiltonian for the quantum spin Hall effect\cite{bernevig2006quantum}.
As a consequence of spin-orbit coupling it features a band inversion around the $\Gamma$ point as illustrated in Fig.~\ref{fig:eh_bands}.
As a combined effect of the time-reversal and inversion symmetries of the system, the conduction and valence bands are each two-fold degenerate.
We label each band by an index denoted as the spin-orbit parity and obtain four eigenstates $\ket{\chi^{\mu}_{\vec{k}}}$, where $\mu \in \{c,v\} \cup \{+,-\}$.

\subsection{Topological exciton eigenstates}

\begin{figure}[!htb]
    \centering
    \includegraphics[width=0.6\linewidth]{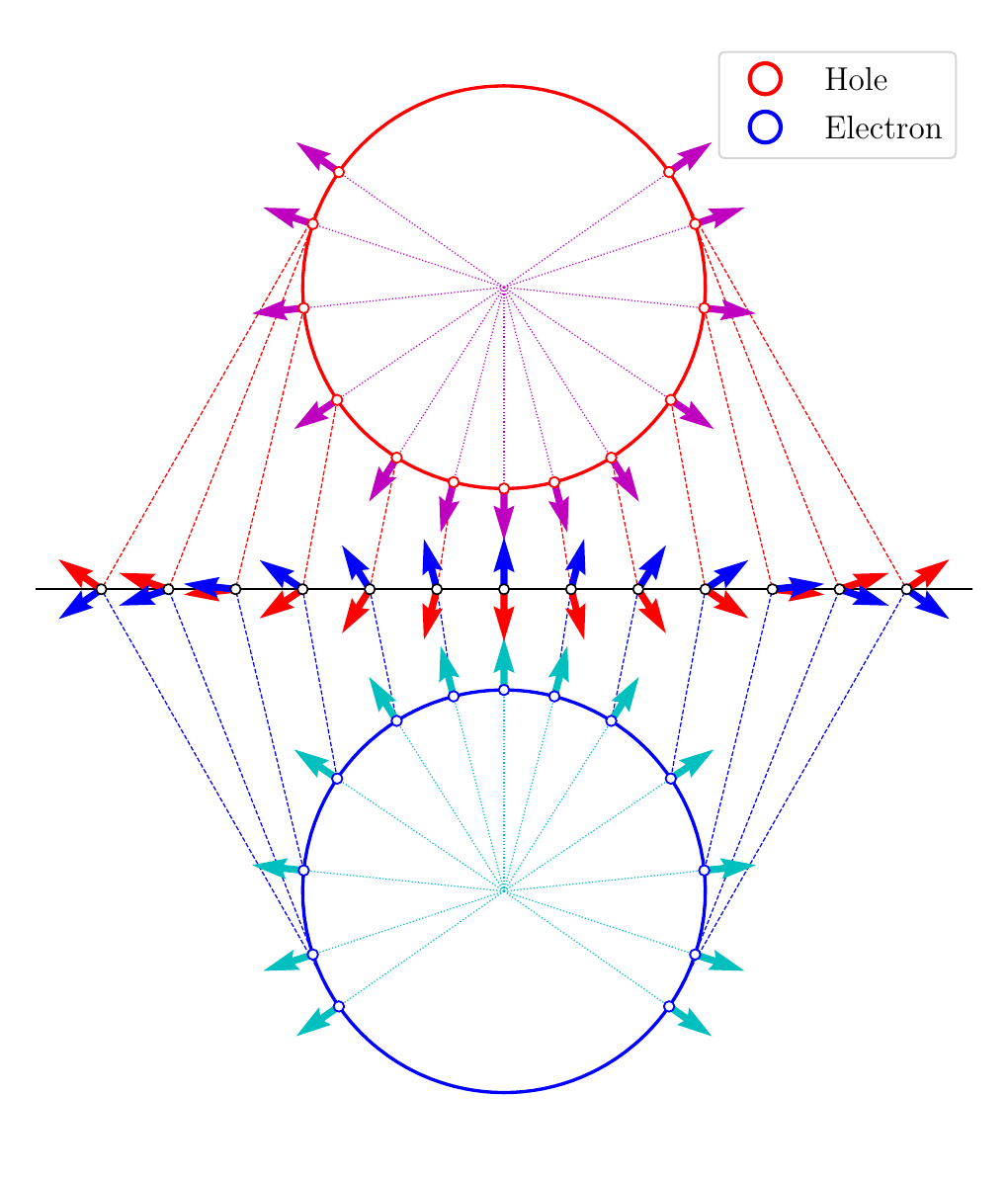}
    \caption{
        \textbf{Momentum-space skyrmion textures of an electron and a hole.}
        This idealized illustration corresponds to the configuration in a weakly bound exciton.
        Due to rotational symmetry the two circles actually represent only a radial slice of the two spheres onto which the complete momentum plane is mapped with a unit winding number.
        The south (north) pole of the top (bottom) sphere corresponds to the origin $Q = 0$, whereas the opposite pole corresponds to $Q \rightarrow \infty$.
        Note that in a normal semiconductor like CdSe the pseudospins always point in the same direction, independent of $Q$, and thus have a zero winding number.
    }
    \label{fig:bloch_sphere_eh}
\end{figure}

The four different spin-orbit-parity combinations for an electron and a hole give four different exciton families $\ket{\vec{Q};s,t}$ that generalize the singlet and triplet states in normal semiconductors.
Here, $s$ and $t$ are the spin-orbit parities of the bands where the electron and the hole are located, respectively, and $\vec{Q} \equiv (Q_{x}, Q_{y})$ is the total in-plane exciton momentum.
In the absence of an exchange interaction, these states diagonalize the two-body Hamiltonian and have a well-defined Chern number equal to $\mathcal{C}_{s,t} = s + t$.
Hence, in this idealized case there are two topologically nontrivial exciton states, characterized by a nontrivial winding of the momentum-space pseudospin texture $\vec{\Gamma}_{e,h}(\vec{Q})$ (defined precisely in the Methods section).
As illustrated in Fig.~\ref{fig:bloch_sphere_eh}, the pseudospin texture gives an intuitive picture of the nontrivial topology of this basis of exciton states, as it allows us to visualize the total Chern number of each basis element as a combination of two winding numbers by looking at the path of the electron (hole) pseudospin from pointing upwards (downwards) at the origin $Q = 0$ to pointing downwards (upwards) at $Q \rightarrow \infty$, represented as the circles in Fig.~\ref{fig:bloch_sphere_eh}.
Thus, the pseudospin texture of each constituent particle can be seen as a skyrmion, so that the topological exciton is represented by a double skyrmion texture.

\begin{figure}[!htb]
    \centering
    \includegraphics[width=0.6\linewidth]{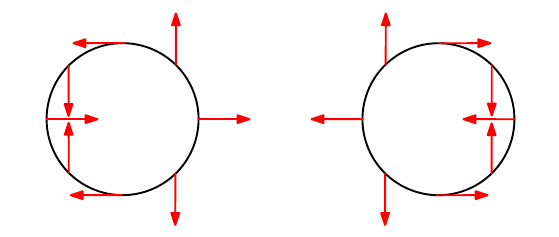}
    \caption{
        \textbf{Behavior of the spin-orbit-parity pseudospin around the origin of momentum.}
        As the polar angle $\phi_{\vec{Q}}$ varies from $0$ to $2 \pi$, the pseudospin $\vec{\sigma}$ of the states $\ket{\vec{Q}; 2_{+}}$ (left) and $\ket{\vec{Q}; 2_{-}}$ (right) winds around twice, as its orientation is coupled to that of the total exciton momentum.}
    \label{fig:pseudospin}
\end{figure}

Including the exchange interaction couples $\ket{\vec{Q};+,+}$ with $\ket{\vec{Q};-,-}$, and this idealized picture complicates somewhat. Analyzing the effective two-dimensional interaction between electrons and holes, we find that the exciton eigenstates split into two doublets as
\begin{align}
    \ket{\vec{Q};0_\pm} &\equiv \frac{1}{\sqrt{2}} \Big( \ket{\vec{Q};+,-} \pm \ket{\vec{Q};-,+} \Big) ~, \\
    \ket{\vec{Q};2_\pm} &\equiv
    \frac{1}{\sqrt{2}} \Big( e^{- i \phi_{\vec{Q}}} \ket{\vec{Q};+,+} \pm e^{i \phi_{\vec{Q}}} \ket{\vec{Q};-,-}
    \Big) ~,
\end{align}
with $\phi_{\vec{Q}}$ the polar angle of $\vec{Q}$ with respect to the $x$-axis. The wave functions contained in $\ket{\vec{Q};+,-}$ and $\ket{\vec{Q};-,+}$ are related by complex conjugation, and so are those contained in $\ket{\vec{Q};+,+}$ and $\ket{\vec{Q};-,-}$. Due to the combined phase factors $e^{\pm i \phi_{\vec{Q}}}$, the members of the doublet $\ket{\vec{Q}; 2_{\pm}}$ have chirality two, a signature of the nonzero Berry curvature within the nanosheet.
This chirality can be understood via the introduction of a spin-orbit-parity pseudospin operator $\vec{\sigma} \equiv (\sigma_{x}, \sigma_{y}, \sigma_{z})$, where $\sigma_{i}$ are the Pauli matrices acting on the space spanned by $\ket{\vec{Q};+,+}$ and $\ket{\vec{Q};-,-}$. Its expected value for the states $\ket{\vec{Q}; 2_{\pm}}$ is
\begin{equation}
   {\langle \vec{\sigma} \rangle}_{2_{\pm}} = \pm \cos(2 \phi_{\vec{Q}}) \, \hat{\vec{x}} \pm \sin(2 \phi_{\vec{Q}}) \, \hat{\vec{y}} ~,
\end{equation}
and its behavior on the momentum plane is sketched in Fig.~\ref{fig:pseudospin}. We see that it has a winding number of $2$ around $\vec{Q} = \vec{0}$, analogously to the valley pseudospin in Ref.~\cite{qiu2015nonanalyticity}.
\textcolor{black}{We mention, in passing, that the chiral excitons considered here are of different nature from those observed in the experiment of Kung et al. \cite{kung2019observation}
Indeed, the latter result from high-energy transitions between massive holes and massless Dirac electrons on the surface, whereas the ones in this work arise from long-wavelength transitions between bulk bands near the Fermi level.}

We may further label the different individual states in $\ket{\vec{Q};0_{\pm}}$ by a principal quantum number $n \in \{0,1,2,\dots\}$ and by their at $\vec{Q} = \vec{0}$ well-defined relative angular momentum $m \in \{-n,\dots,0,\dots,n\}$, introducing the notation $\ket{\vec{Q};0_{\pm};n, m}$.
On the other hand, the individual states in $\ket{\vec{Q};2_{\pm}}$ are similarly labeled by the at $\vec{Q} = \vec{0}$ well-defined angular momentum $m$ of the $\ket{\vec{Q};+,+}$ component in their linear combination.

\subsection{Exciton dispersion relations and wave functions}


\begin{figure}[!htb]
    \begin{center}
       \includegraphics[width=0.6\linewidth]{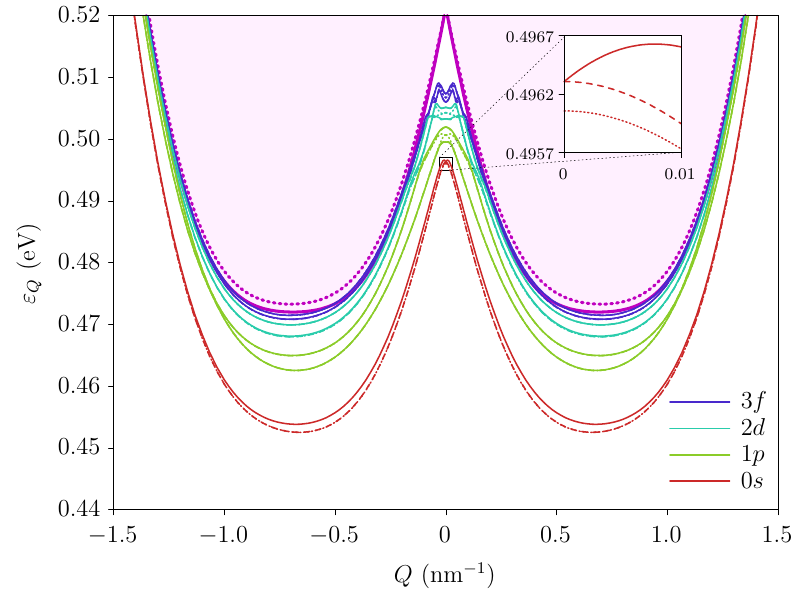}
       \caption{
       \textbf{Excitonic dispersion relation for the \HN potential.}
       Shown are the exciton eigenstates $\ket{\vec{Q}; 0_{\pm}}$ (dotted lines), $\ket{\vec{Q};2_{+}}$ (solid lines), and $\ket{\vec{Q};2_{-}}$ (dashed lines), with $\vec{Q}$ the total exciton momentum.
       Different colors correspond to different values of $m$ at $\vec{Q} = \vec{0}$ as indicated in the legend.
       The $1s$ state (not shown) has higher energy than the $1p$, $2d$ and $3f$ states.
       \textcolor{black}{The excitons $\ket{\vec{Q},2_{\pm}}$ with opposite angular momenta are split in energy, with the higher state having angular momentum $m > 0$ and the lower state $m < 0$.}
       The magenta region at the top represents the electron-hole continuum, delimited by a solid purple line resulting from analytically minimizing the energy gap with respect to the relative electron-hole momentum in the absence of interactions.
       The purple dotted line represents the continuum threshold as calculated numerically including only angular momenta $|m| \leq 3$ and tends to the solid line upon inclusion of higher values of $|m|$.}
       \label{fig:exc_disp}
    \end{center}
\end{figure}

\begin{figure}[!htb]
    \begin{center}
       \includegraphics[width=0.6\linewidth]{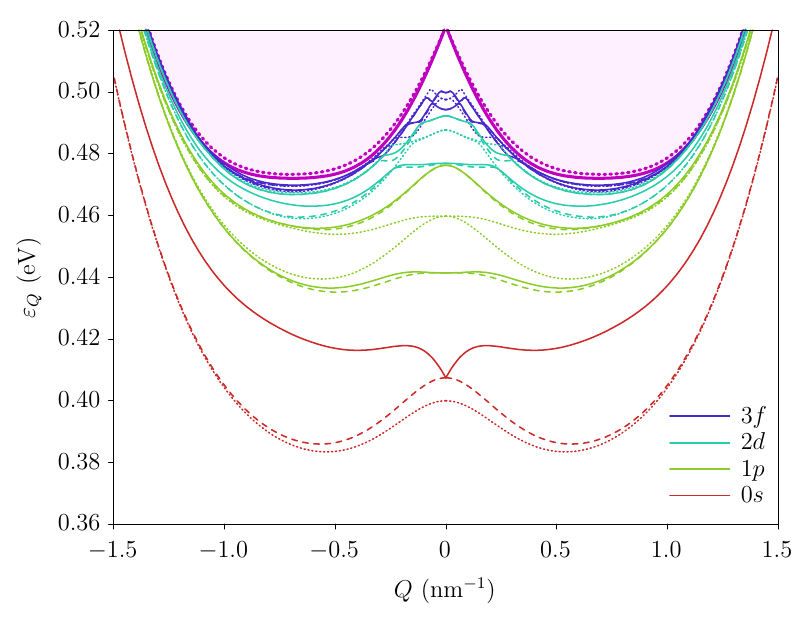}
       \caption{\textbf{Excitonic dispersion relation for the Coulomb potential.}
       All colors and line types are equivalent to their counterparts in Fig.~\ref{fig:exc_disp}.}
       \label{fig:exc_disp_C}
    \end{center}
\end{figure}

We have solved the associated Bethe-Salpeter equation with both the \HN potential and the two-dimensional Coulomb potential after neglecting the effects of the quantum confinement in the $z$-direction.
The former is often used to approximate the long-distance behavior of the microscopic Keldysh potential\cite{keldysh1979}.
We find that only the Rytova-Keldysh interaction is compatible with the neglect of quantum confinement, so unless otherwise specified all results correspond to those obtained with this potential.
Figs.~\ref{fig:exc_disp} shows the dispersion relations of all four excitonic ground states, and also of several excited states that have a nonzero angular momentum at $\vec{Q} = \vec{0}$.
The same is shown in Fig.~\ref{fig:exc_disp_C} for the Coulomb interaction, whose weaker nature allows us to better visualize the features of the band structure that are qualitatively independent of the interaction details due to the robustness of the topology.
Notice the difference in behavior of the energy between the $\ket{\vec{Q};2_\pm;0,0}$ and $\ket{\vec{Q};0_\pm;0,0}$ doublets around zero momentum, as the former linearly splits off.
This effect is heavily influenced by the topology, as we have checked that the dispersions become strongly parabolic in the absence of a band inversion, \textcolor{black}{with the nonanalytic mode being barely noticeable for all states}.
The linear dispersion may be analytically understood by expanding the effective exchange potential around this point, which is further analyzed in Ref.~\cite{andreani1990}. Moreover, the effective $2 \times 2$ Hamiltonian for the $\ket{\vec{Q};2_\pm;0,0}$ chiral doublet has been discussed previously \cite{yu2014a,wu2015a,qiu2015nonanalyticity} and is further studied below.
The energy of the $\ket{\vec{Q};0_\pm;0,0}$ doublet follows a quadratic dispersion both at small and large momenta, with a crossover taking place around the minimum of the electron-hole continuum, as the inversion of the bands after that has less of an effect.

Interestingly, the states $\ket{\vec{Q};2_{\pm}}$ with opposite angular momenta $+m$ and $-m$ are split in energy, with the former having higher energy than that of the latter, as seen in the figures for the solid and dashed lines.
This is a consequence of the Berry curvature within the nanosheet\cite{zhou2015berry}, which provides an anomalous contribution to the single-particle velocity that is only present in the subspace spanned by the $\ket{\vec{Q};+,+}$ and $\ket{\vec{Q};-,-}$ states.
\textcolor{black}{Our results agree qualitatively with previous works that perturbatively incorporate the effects of the Berry curvature \cite{hichri2019,trushin2018,zhou2015berry,srivastava2015a,sablikov2017,allocca2018}.
However, numerical discrepancies are expected between these references and the present work, as the former all consider the effective-mass approximation.
This is not appropriate in our system due to the band inversion, which prevents us from decoupling the relative and center-of-mass motions.
Consequently, our exciton spectra at zero momentum deviate from the well-known Rydberg series.}

Also, the energies of the higher excited states closely follow the electron-hole continuum, as the excitons are more delocalized in real space and their wave functions become those of a separate electron-hole pair.
An important feature of the \HN spectrum is that all excitons are strongly indirect, which leads to long lifetimes due to the strongly reduced radiative recombination rate.
As revealed by the shape of the electron-hole continuum, this is a direct consequence of the band inversion of the underlying single-particle bands.
Note, however, that in this work both the electron and the hole are taken to reside in the same layer and are thus still direct in real space.

\begin{figure}[!htb]
    \begin{center}
        \includegraphics[width=0.6\linewidth]{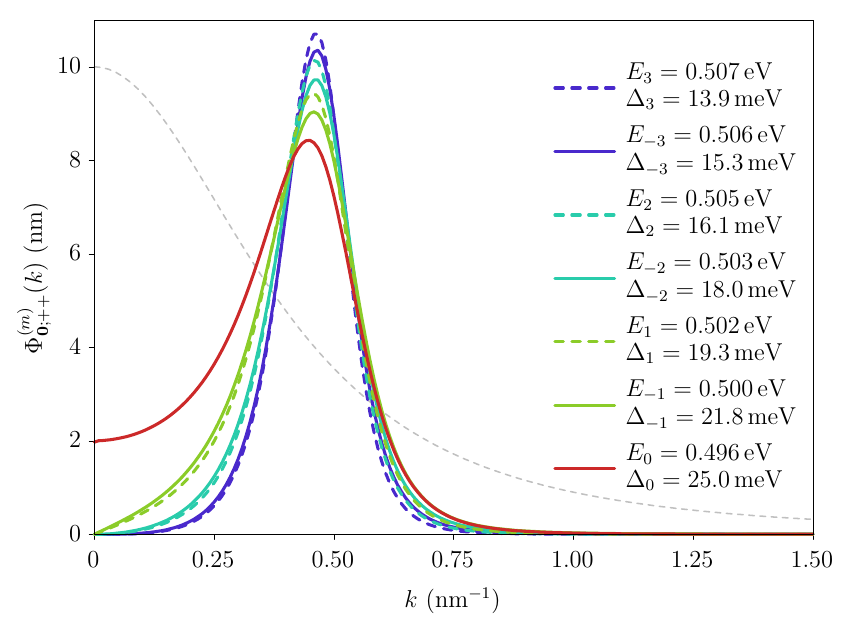}
        \caption{
            \textbf{Magnitudes of momentum-space exciton wave functions.}
            These correspond to the zero-momentum states $\ket{\vec{0};+,+;n,m}$ and $\ket{\vec{0};-,-;n,-m}$ and are shown for several values of $m$ as ordered in Fig.~\ref{fig:exc_disp} and the first available $n$.
            The wave functions themselves are obtained by multiplying the magnitude by the corresponding phase $e^{i m \phi_{\vec{k}}}$.
            The wave functions have a maximum around the momentum at which the energy gap presents a minimum.
            In particular, the wave function in the $s$-wave case (red line) significantly differs from that obtained from the hydrogen problem, which is proportional to $((a_0 k)^2 + 1)^{-3/2}$ (grey dashed line). Here we have set $a_{0} = 10/\sqrt{8 \pi}\,\si{\nano\meter} \simeq \SI{1.99}{\nano\meter}$, which is suitable for comparison.
            The values $E_m$ shown in the figure correspond to the excitonic eigenenergies for the same states.
            The quantities $\Delta_m$ are the binding energies of each state, that is, the difference between the electron-hole continuum and $E_m$.
        }
        \label{fig:exc_wave function}
    \end{center}
\end{figure}

\begin{figure}[!htb]
    \begin{center}
        \includegraphics[width=0.5\linewidth]{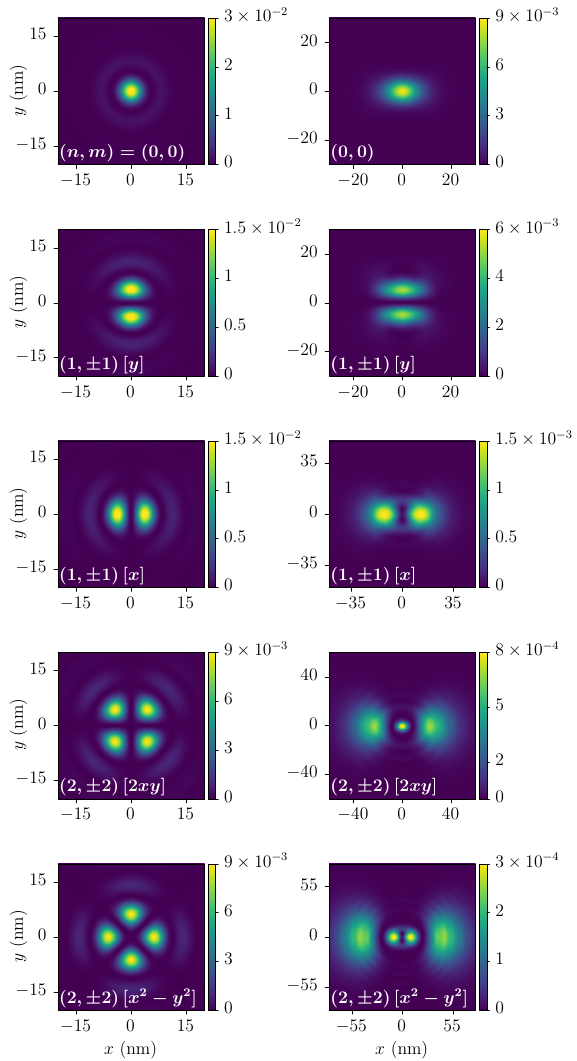}
        \caption{
        \textbf{Relative real-space probability densities of the exciton eigenstates $\vec{\ket{\vec{Q}; 0_{\pm}; n, m}}$.}
        The rows correspond, from top to bottom, to the first five dotted-line states of Fig.~\ref{fig:exc_disp}.
        In the left column, $\vec{Q} = \vec{0}$ and we take appropriate linear combinations resulting in hydrogen-like orbitals.
        The right column corresponds to $\vec{Q}$ along the $x$-direction, with $Q = \SI{0.7}{\nano\meter\tothe{-1}}$.
        \textcolor{black}{The values of $n$ and $m$ are indicated in each plot, as well as the orbital name in the standard notation.}
        }
        \label{fig:PDUncoupled}
    \end{center}
\end{figure}

\begin{figure}[!htb]
    \begin{center}
        \includegraphics[width=0.696\linewidth]{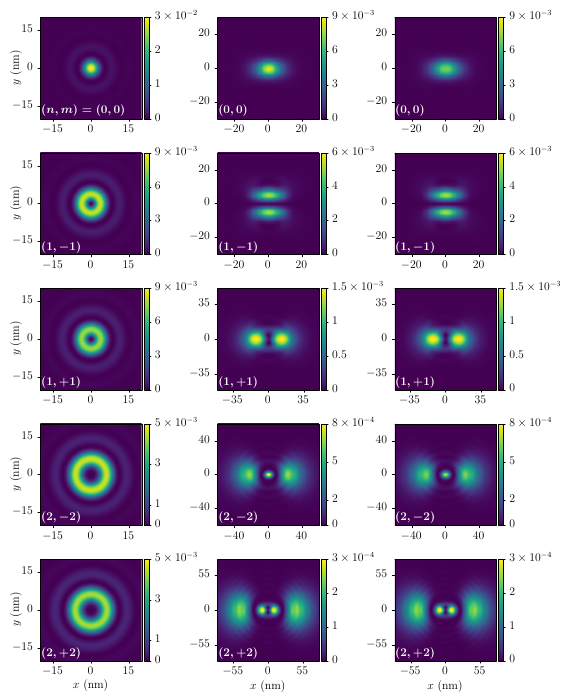}
        \caption{
        \textbf{Relative real-space probability densities of the exciton eigenstates $\vec{\ket{\vec{Q}; 2_{\pm}; n, m}}$.}
        The rows correspond, from top to bottom, to the first five solid- and dashed-line states of Fig.~\ref{fig:exc_disp}.
        More precisely, the states shown are $\ket{\vec{0}; 2_{\pm}; n, m}$ (left column), $\ket{\vec{Q}; 2_{-}; n, m}$ (middle column), and $\ket{\vec{Q}; 2_{+}; n, m}$ (right column).
        \textcolor{black}{The values of $n$ and $m$ are indicated in each plot.}
        We have set $\vec{Q}$ along the $x$-direction with $Q = \SI{0.7}{\nano\meter\tothe{-1}}$.
        }
        \label{fig:PDCoupled}
    \end{center}
\end{figure}

Fig.~\ref{fig:exc_wave function} shows the relative exciton wave function of the states $\ket{\vec{Q};2_{\pm}}$ for the ground state and several excited states from Fig.~\ref{fig:exc_disp} at $\vec{Q} = \vec{0}$.
Note the significantly different behavior of the ground-state wave function compared to that obtained from the hydrogen problem, which is proportional to ${((a_0 k)^2 + 1)^{-3/2}}$.
Furthermore, Figs.~\ref{fig:PDUncoupled} and \ref{fig:PDCoupled} show the relative real-space probability density of the low-lying states $\ket{\vec{Q}; 0_{\pm}}$ and $\ket{\vec{Q}; 2_{\pm}}$, respectively. For the states $\ket{\vec{Q};0_{\pm}}$ at $\vec{Q} = \vec{0}$ we exploit their opposite-angular-momentum degeneracy to obtain linear superpositions resulting in hydrogen-like orbitals shown in the left column, and note how these become deformed in the direction of a nonzero exciton momentum in the right column due to the breaking of the rotational symmetry. In the case of the states $\ket{\vec{Q}; 2_{\pm}}$, the splitting between states with opposite $m$ at $\vec{Q} = \vec{0}$ prevents us from writing down such orbitals, so the first column shows a rotationally invariant probability density. Nevertheless, a nonzero exciton momentum breaks again this rotational symmetry, and the wave functions develop lobes in either the transversal or the longitudinal direction as seen in the second and third columns.

\subsection{Optical properties}

We have also derived selection rules for circularly polarized light in the $xy$-plane moving in the positive $z$-direction at $\vec{Q} = \vec{0}$, where the exchange interaction vanishes and the exciton families $\ket{\vec{Q};+,+}$ and $\ket{\vec{Q};-,-}$ become uncoupled.
For left-circularly polarized light (with angular momentum ${m_{\gamma} = +1}$) we find that the excitons $\ket{\vec{0}; +,+; n, 0}$ and $\ket{\vec{0}; -,-; n, {+}2}$ are bright, whereas the rest are dark.
On the other hand, for right-circularly polarized light ${(m_{\gamma} = {-}1)}$ the only optically active excitons are $\ket{\vec{0}; -,-; n, 0}$ and $\ket{\vec{0}; +,+; n, {-}2}$. Note that these results combined are in accord with the time-reversal symmetry.
By contrast, the excitons $\ket{\vec{0};+,-}$ and $\ket{\vec{0};-,+}$ are all dark irrespective of their angular momentum.
These results greatly differ from the situation in ordinary semiconductors, where only the $s$-wave singlet is bright.

\subsection{Topological excitonic edge states}\label{sec:5}

For small momenta, the behavior of each pair of $\ket{\vec{Q};+,+}$ and $\ket{\vec{Q};-,-}$ excitons can be understood by means of an effective $2 \times 2$ model.
Here we are interested in the effects of their topological properties, and in particular we focus on the $s$-wave ground state.
For our purposes it is enough to study the minimal Hamiltonian $H^{\mathrm{eff}}(\vec{Q}) = \vec{b}(\vec{Q}) \vec{\cdot} \vec{\sigma}$, with $\vec{b}(\vec{Q}) = (J Q \cos 2 \phi_{\vec{Q}}, \, J Q \sin 2 \phi_{\vec{Q}}, \, \Delta)$.
Here, $J$ is the exchange coupling and $\vec{\sigma}$ the spin-orbit-parity pseudospin, and we have added a possible Zeeman-like term $\Delta \sigma_{z}$ that breaks the time-reversal symmetry.
The topological properties become apparent after calculating the winding number of the pseudo-magnetic field $\vec{b}(\vec{Q})$, which is $w = \operatorname{sgn} \Delta$.
This unit winding when $\Delta \neq 0$ can be understood from the fact that the normalized vector $\hat{\vec{b}}(\vec{Q})$ represents a meron configuration with chirality two.
Even though we have neglected higher-order terms in the effective Hamiltonian, exciton edge states can in principle be found with these terms included as well, as argued in the Methods section.
The approximation performed here admits a simple analytic solution, whereas the expressions become very cumbersome when quadratic terms are kept.

\begin{figure}[!htb]
    \centering
    \includegraphics[width=0.6\linewidth]{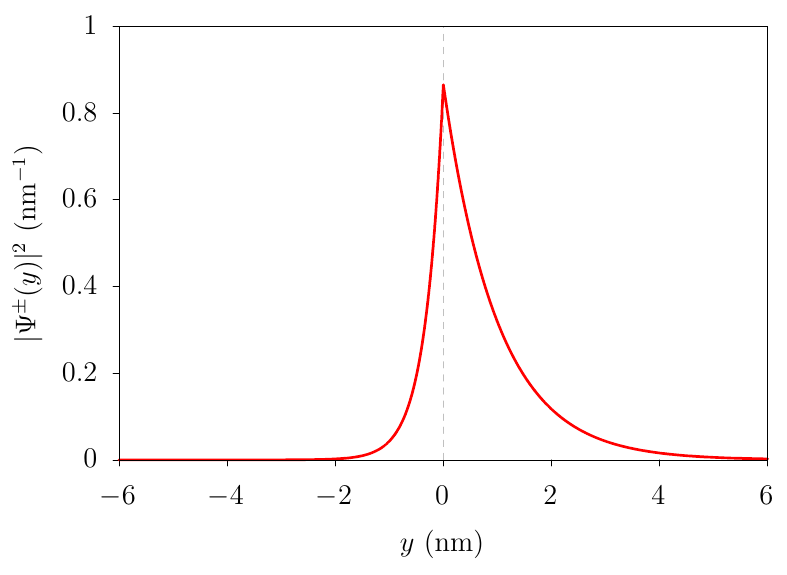}
    \caption{
    \textbf{Probability density of chiral exciton edge states.}
    Here, $Q_{x} = 0$, and we have used the parameters $\Delta_{1} = 3 \Delta_{2}$ and $|\Delta_{2}/J| = 1$.
    The states are localized at the interface $y = 0$ and their wave functions decay exponentially with the distance from this point.}
    \label{fig:edgeStates}
\end{figure}

We consider the setting of two semi-infinite systems defined by $y < 0$ and $y > 0$, with Zeeman-like couplings $\Delta_{1}$ and $\Delta_{2}$, respectively. We have solved the Hamiltonian for $Q_{x} = 0$, in which case there are two zero-energy eigenstates that read
\begin{equation}\label{eq:WFedge}
    \Psi^{\pm}(y) = \sqrt{\frac{1}{|J|} \frac{|\Delta_{1}||\Delta_{2}|}{|\Delta_{1}| + |\Delta_{2}|}} \begin{bmatrix}
        1 \\ \pm i \eta \operatorname{sgn} J
    \end{bmatrix} \exp\bigg({-}\bigg|\frac{\Delta(y)}{J} y\bigg|\bigg) ~,
\end{equation}
where $\eta \equiv {-}\operatorname{sgn} \Delta_{1} = \operatorname{sgn} \Delta_{2}$.
These solutions describe states that are localized around the interface at $y = 0$ and decay exponentially away from this point.
We stress that they only exist when $\operatorname{sgn} \Delta_{1} = {-}\operatorname{sgn} \Delta_{2}$, as there is no continuous normalizable solution when the signs of these Zeeman-like couplings are equal.
Furthermore, these states are robust against perturbations of the gap parameter at either side, and in particular the solution exists even in the limit ${|\Delta_{1}| \rightarrow \infty}$, even though in this case the continuity condition must be relaxed.
In Fig.~\ref{fig:edgeStates} we have plotted a typical example of the corresponding probability density $|\Psi^{\pm}(y)|^{2}$.

\textcolor{black}{We briefly comment on the term $\Delta \sigma_{z}$, as it has been introduced by hand in the effective Hamiltonian.
This is necessary because the underlying BHZ model is time-reversal symmetric and thus by itself will not give rise to such a term.
However, the topological properties only depend on its sign, not on its magnitude, and emerge no matter how small $\Delta$ may be as long as it is nonzero. We are therefore allowed to add it to the effective model as a perturbation.
Experimentally, this term may be realized by a time-reversal-breaking perturbation such as a small magnetic field, via contact to a thin magnetized layer, or with the injection of a small concentration of magnetic impurities.
Furthermore, even though the topological properties imbued by $\Delta$ may seem independent of the state of the underlying BHZ Hamiltonian, one must keep in mind that the effective model for the chiral excitons arises only in the topological regime.
Consequently, both ingredients are required for the emergence of topological excitons.}

For small but nonzero $Q_{x}$, perturbation theory yields an equal energy shift for both edge states, namely
\begin{equation}\label{eq:energyShift}
    E_{\pm}(Q_{x}) = {-} 2 \eta |J| Q_{x} + \mathcal{O}(Q_{x}^{2}) ~.
\end{equation}
Thus, the system hosts two chiral modes of excitons at the interface, with a Fermi velocity of magnitude $2|J|$ and with both states moving in the same direction.
Due to the factor of $\eta$ in the dispersion relation, this direction is directly specified by the bulk winding number at either side of the boundary.
Hence, reversing the Zeeman-like coupling at both sides also reverses the direction of the chiral excitons, as expected from the analogy of $\Delta$ with a magnetic field.
These results can be directly understood from the bulk-boundary correspondence, which predicts a total number of states equal to the difference in winding numbers at both sides, $|w_{y < 0} - w_{y > 0}| = 2$, with their direction being determined by the sign of this difference \cite{mong2011edge}.

\section{Discussion}

Our results show that, in principle, the topology of the conduction and valence bands is indeed inherited by the Wannier excitons, as the exciton wave function contains electron and hole pseudospin textures with a topologically nontrivial winding. This is in particular true for the excitonic basis states that diagonalize the Wannier problem with only the direct interaction included. As the exchange interaction couples these states the physical picture complicates somewhat, but the nontrivial pseudospin winding remains. Furthermore, the bare two-dimensional interaction is heavily modified by the topological band structure. For the excitonic ground state, this ultimately results in a nonchiral doublet with quadratic dispersion relation at low momenta, and a chiral doublet with one linearly dispersing mode and one quadratic mode. Diagonalization of the resulting low-momentum effective model including a Zeeman-like coupling leads to topological edge states of chiral excitons. Since our work does not impose many restrictions on the model parameters, we expect that it remains valid for other similar topological materials such as Bi\textsubscript{2}Te\textsubscript{3} and Sb\textsubscript{2}Se\textsubscript{3} by using the appropriate parameter values.

\textcolor{black}{A future direction of research is to consider the motion of the excitons under the effects of lattice strain, which yields a confining potential in real space.
In this scenario we expect that the nontrivial Berry curvature, which constitutes a momentum-space magnetic field, will give rise to an anomalous Hall effect\cite{cao2021quantum,chaudhary2021anomalous}.}
Experimentally, we want to resolve the linear dispersion in the chiral doublet, which would be a first step towards the observation of the topological properties of excitons.
This is in principle possible by means of terahertz spectroscopy, as the details of the dispersion affect the chemical equilibrium between excitons and free charges in such pump-probe experiments \cite{garciaflorez2019,lauth2017ultrafast,lauth2019}.
We also expect the polarizability of the excitons to be strongly affected by their topology, which can again be observed in terahertz conductivity measurements.
In particular, the polarizability of the topological $s$-wave excitons should be significantly reduced with respect to that of their trivial counterparts due to their modified relative wave function.
Another interesting feature of the obtained excitonic band structure is that it is indirect as a consequence of the band inversion of the single-particle Hamiltonian, which leads to a greatly reduced rate of radiative recombination processes and thus to long-lived excitons.

\textcolor{black}{The selection rules we have derived can be understood by noting that the eigenstates $\ket{\chi^{\mu}_{\vec{k}}}$ are also eigenstates of the total angular momentum operator $j_{z} = l_{z} + \frac{1}{2} s_{z}$, where $s_{z}$ is the Pauli matrix acting on the spin part of the single-particle basis and $l_{z} = -i \partial_{\phi_{\vec{k}}}$.}
Consequently, the single-particle states $\ket{\chi^{c,s}_{\vec{k}}}$ and $\ket{\chi^{v,t}_{\vec{k}}}$ have angular momenta $\frac{s}{2}$ and ${-}\frac{t}{2}$, respectively.
By noting that the angular momentum of a hole is opposite to that of the destroyed valence electron, one finds that an exciton can couple to a left-circular photon if its relative angular momentum is $0$ and the underlying electron and hole lie in the bands with $s = t = +1$, or if its relative angular momentum is $+2$ and the particles lie in the bands with $s = t = -1$, as we have seen.
The argument follows analogously for right-circular photons.

Finally, the (top and bottom) surface electronic states, which have been neglected in this work, may contribute to several aspects that will be analyzed in a follow-up article\cite{maisel2022}.
First, by screening the interaction between bulk electrons and holes, which may be treated as static screening of the Coulomb or \HN potentials.
We have in first instance neglected this effect because of the reduced density of states around the Dirac cone, to which the screening length will be inversely proportional, at least in the Thomas-Fermi regime.
Note that we are always considering single-exciton excitations in the material, so that electron-electron correlation effects are effectively incorporated in the parameters of the band structure, for instance via GW corrections.
Second, by causing a nonzero transition probability from bulk states to surface states mediated by surface plasmons, which thus in principle provide a mechanism for surface electron-hole decay of bulk excitons.
The effects of both the screening and exciton decay will be analyzed perturbatively to verify the correctness of our results, as it depends on the magnitude of the energy shifts and spectral broadenings induced by these phenomena.
\textcolor{black}{Away from zero chemical potential, we expect quasi-2D nanosheets of Bi\textsubscript{2}Se\textsubscript{3} with around 6 QLs to provide a promising platform for exciton-plasmonics, which will constitute the main topic of the sequel.}
Third, by a contribution to the complex conductivity measured in pump-probe terahertz conductivity experiments, which may be modelled as a separate contribution from that of the bulk states, as similarly done for unbound charges in Ref.~\cite{garciaflorez2019}.
However, we expect the contribution from the surface states to the absorption to be small compared to that of the bulk states, as the oscillator strength of the latter is much larger than that of the former.

\section{Methods}

\noindent\subsection{Band structure of Bi\textsubscript{2}Se\textsubscript{3} nanosheets}

Our starting point is the $\vec{k}\boldsymbol{\cdot}\vec{p}$ Hamiltonian derived in \cite{zhang2009} and \cite{liu2010} for modeling three-dimensional Bi\textsubscript{2}Se\textsubscript{3} and Bi\textsubscript{2}Te\textsubscript{3} around the $\Gamma$ point in the Brillouin zone. To account for the quantization in the $z$-direction in a nanosheet geometry of thickness $L_{z}$, we solve the model at the two-dimensional $\Gamma$ point ${(k_{x} = k_{y} = 0)}$ with the substitution ${k_{z} \rightarrow -i \partial_{z}}$ and hard-wall boundary conditions as done by \cite{lu2010} and \cite{shan2010effective}. Everywhere in our numerics we consider ${L_{z} = \SI{6}{\nano\meter}}$, which in the case of Bi\textsubscript{2}Se\textsubscript{3} is sufficiently large for the single-particle Dirac states on the opposite surfaces not to be gapped out by tunneling processes and the nontrivial topology of the bulk to survive \cite{zhang2009quintuple,zhang2010crossover,taskin2012manifestation}, but also sufficiently small for bulk electrons and holes to still behave as in two dimensions. We then project the 3D Hamiltonian onto the energetically highest-lying valence and lowest-lying conduction subbands in order to integrate out the $z$-dependence. Hence, we assume that the relevant low-energy physics of individual particles is confined to this subspace, which has been verified {\it a posteriori} by comparing the obtained exciton binding energies to the energy splitting of the bulk subbands due to the confinement in the $z$-direction.

Ultimately, after projecting the three-dimensional $\vec{k}\vec{\cdot}\vec{p}$ Hamiltonian onto the single-particle ground states of the 3D model, we obtain our desired nanosheet Hamiltonian. Written in terms of Pauli matrices in spin and orbital space, $\vec{s}$ and $\vec{\tau}$ respectively, it is given by the $4\times4$ matrix
\begin{equation}\label{eq:HamiltonianProjected}
    H_0(\vec{k}) = \epsilon_0(\vec{k}) + M(\vec{k})\tau_z + A_2(k_x s_x + k_y s_y) \tau_x
    ~ ,
\end{equation}
where $\vec{k} \equiv (k_{x}, k_{y})$ is the in-plane momentum, $\epsilon_0(\vec{k}) \equiv E + D_2 (k_x^2 + k_y^2)$, and $M(\vec{k}) \equiv M - B_2 (k_x^2 + k_y^2)$.
Note that products of matrices in different spaces, e.g., $s_x \tau_x$, are Kronecker products and not matrix products, and that identity matrices are implied.
Our effective two-dimensional Hamiltonian is equivalent to that of the BHZ model after a suitable unitary transformation.
Furthermore, as expressed in Eq.~\eqref{eq:HamiltonianProjected} it has the same form as the three-dimensional Bi\textsubscript{2}Se\textsubscript{3} Hamiltonian with $k_{z} = 0$ and some renormalized values of the parameters with respect to those given by \cite{zhang2009}
However, the basis in which it is expressed differs from that of the 3D model, since the original $\mathrm{Bi}^{+}$ and $\mathrm{Se}^{-}$ orbitals have hybridized in the corresponding eigenstates.
We denote these hybridized orbitals by $\mathrm{Bi'}^{+}$ and $\mathrm{Se'}^{-}$.
The renormalized values of the parameters, with which all of our numerical results have been obtained, are $A_2 = \SI{0.41}{\electronvolt\nano\meter}$, $M = \SI{0.28}{\electronvolt}$, $B_2 = \SI{0.473}{\electronvolt\nano\meter\squared}$, $E = \SI{-0.0012}{\electronvolt}$, and $D_2 = \SI{0.202}{\electronvolt\nano\meter\squared}$.

Diagonalization of $H_{0}(\vec{k})$ leads to the conduction and valence band energies
\begin{equation} \label{eq:eh_bands}
    \epsilon_{c,v}(\vec{k}) = \epsilon_0(\vec{k}) \pm \sqrt{M(\vec{k})^2 + A_2^2(k_x^2+k_y^2)} ~,
\end{equation}
as well as to the topologically nontrivial single-particle states that we use throughout in our description of excitons.
As a consequence of time-reversal symmetry in combination with inversion symmetry, the Hamiltonian does not couple the two subspaces $\{ \ket{\mathrm{Bi'}^{+};\uparrow},\ket{\mathrm{Se'}^{-};\downarrow} \}$ and $\{ \ket{\mathrm{Bi'}^{+};\downarrow},\ket{\mathrm{Se'}^{-};\uparrow} \}$, and the conduction and valence bands are each two-fold degenerate.
The corresponding eigenstates are labeled by their so-called spin-orbit parity, defined as the eigenvalue of the operator $s_{z} \tau_{z}$ that commutes with $H_{0}(\vec{k})$.

We next introduce the vector of pseudospin operators $\vec{\Gamma} \equiv (s_x\tau_x,\,s_y\tau_x,\,\tau_z)$, which are used to rewrite Eq.~\eqref{eq:HamiltonianProjected} as $H_0(\vec{k}) = \epsilon_0(\vec{k}) + \vec{d}(\vec{k}) \boldsymbol{\cdot} \vec{\Gamma}$, with $\vec{d}(\vec{k}) \equiv (A_2 k_x, \, A_2 k_y, \, M(\vec{k}))$.
The nontrivial topology of the Hamiltonian is now very explicit in this form, since $\vec{d}(\vec{k})$ is a skyrmion texture in the momentum plane $(k_x, k_y)$ and the operator $\vec{\Gamma}$ reduces exactly to the three Pauli matrices in the uncoupled even and odd spin-orbit-parity subspaces.
Therefore, the expectation value of the pseudospin as a function of $\vec{k}$ follows the nontrivial winding of $\vec{d}(\vec{k})$ and the winding number of the latter is up to a possible sign equal to the Chern number of the conduction and valence bands.

\subsection{Exciton basis and wave functions}

The conduction and valence states are $\braket{\vec{x};a}{\vec{k};\mu} = (e^{i \vec{k}\boldsymbol{\cdot}\vec{x}}/\sqrt{V}) \braket{a}{\chi_{\vec{k}}^{\mu}}$, where $V = L_x L_y$, the states $\ket{a}$ denote our four-dimensional combined spin and orbital basis states, and $\ket{\chi^{\mu}_{\vec{k}}}$ with $\mu \in \{c,\pm;v,\pm\}$ are the eigenstates of the Hamiltonian in Eq. \eqref{eq:HamiltonianProjected}. Note that the latter are not periodic due to the fact that our Hamiltonian is derived from $\vec{k} \vec{\cdot} \vec{p}$ theory and only accurately models the band structure around the $\Gamma$ point.
An exciton state $X \equiv \{\vec{Q},s,t\}$ is labeled by the total momentum $\vec{Q}$ and the spin-orbit-parities $s$ and $t$ of the electron and hole bands, respectively, as well as by an additional set of ro-vibrational quantum numbers that we introduce later. Such an exciton state is a bound state in the polarization \cite{haug2009quantum}, which in second quantization is determined by the pair correlation function
\begin{align} \label{eq:cv_sec_q}
    \big\langle\hat\psi_{a\vphantom{tb}}^{\vphantom{\dagger}}&(\vec{x}_1) \hat\psi_{b}^\dagger(\vec{x}_2)\big\rangle_{X} \notag \\
    &\hspace{-4mm} = \sum_{\vec{k}} \Phi^{s,t}_{\vec{Q},\vec{k}} \braket{\vec{x}_1;a}{\vec{Q}/2 + \vec{k};c,{s}} \braket{-\vec{Q}/2 + \vec{k}; v,{t}}{\vec{x}_2;b} \notag \\
    &\hspace{-4mm}= \frac{e^{i \vec{Q} \vec{\cdot} \vec{R}}}{V} \sum_{\vec{k}} \Phi^{s,t}_{\vec{Q},\vec{k}} \, e^{i \vec{k}\boldsymbol{\cdot} (\vec{x}_1 - \vec{x}_2)} \braket{a}{\chi_{\vec{Q}/2 + \vec{k}}^{c,{s}}} \braket{\chi_{{-}\vec{Q}/2 + \vec{k}}^{v,{t}}}{b} ~,
\end{align}
where $s$, $t \in \{+,-\}$ and $\vec{R} \equiv (\vec{x}_{1} + \vec{x}_{2})/2$ is the position of the exciton.
Here $\Phi^{s,t}_{\vec{Q},\vec{k}}$ is the relative wave function of the exciton in momentum space.
\textcolor{black}{One may also perform a particle-hole transformation so that holes become positive-energy excitations.
In terms of electrons and holes, the conduction states remain unchanged, $\braket{a}{\chi_{\vec{q}}^{e,s}} = \braket{a}{\chi_{\vec{q}}^{c,s}}$.
By contrast, the hole states satisfy $\braket{b}{\chi_{-\vec{q}}^{h,t}} = \braket{\chi_{\vec{q}}^{\vphantom{h}v,t}}{b}$.}
In this picture the pair correlation function thus reads
\begin{align}\label{eq:eh_exc_sec_q}
    \big\langle\hat\psi_{a}&(\vec{x}_1)\hat\psi_{b}(\vec{x}_2)\big\rangle_{X} \notag \\
    &\hspace{-4mm} = \frac{e^{i \vec{Q} \vec{\cdot} \vec{R}}}{V} \sum_{\vec{k}} \Phi^{s,t}_{\vec{Q},\vec{k}} \, e^{i \vec{k}\boldsymbol{\cdot} (\vec{x}_1 - \vec{x}_2)} \braket{a}{\chi_{\vec{Q}/2 + \vec{k}}^{e,{s}}} \braket{b}{\chi_{\vec{Q}/2 - \vec{k}}^{h,{t}}}~.
\end{align}
If desired we can then also obtain a first-quantized wave function, which is given by
\begin{align} \label{eq:exc_first_q}
    \Psi^{X}_{ab}(\vec{x}_1, \vec{x}_2) = &\frac{1}{V} \frac{e^{i \vec{Q} \vec{\cdot} \vec{R}}}{\sqrt{V}} \sum_{\vec{k}} e^{i \vec{k}\boldsymbol{\cdot} (\vec{x}_1 - \vec{x}_2)} \notag \\
    &\times \frac{1}{\sqrt{2}} \left(
    \Phi^{s,t}_{\vec{Q},\vec{k}} {\braket{a}{\chi_{\vec{Q}/2+\vec{k}}^{\vphantom{h}e,{s}}}}_{1} {\braket{b}{\chi_{\vec{Q}/2-\vec{k}}^{h,{t}}}}_{2} - \Phi^{s,t}_{\vec{Q},{-}\vec{k}} {\braket{b}{\chi_{\vec{Q}/2-\vec{k}}^{\vphantom{h}e,{s}}}}_{2} {\braket{a}{\chi_{\vec{Q}/2+\vec{k}}^{h,{t}}}}_{1} \right)     ~.
\end{align}
This is normalized provided that $\frac{1}{V} \sum_{\vec{k}} |\Phi^{s,t}_{\vec{Q},\vec{k}}|^{2} = 1$. Note that interchanging both particles, thus $\vec{x}_{1} \leftrightarrow \vec{x}_{2}$ and $(a,1) \leftrightarrow (b,2)$, results in an overall minus sign after we perform the transformation $\vec{k} \rightarrow -\vec{k}$.

The single-particle states, together with the relative momentum states ${\ket{\vec{k}}}_{\mathrm{rel}}$, describe the exciton states $\ket{\vec{Q};s,t}$ fully in Dirac notation as
\begin{equation}\label{eq:exc_state_Q}
    \ket{\vec{Q};s,t} = \frac{1}{\sqrt{V}} \ket{\vec{Q}} \sum_{\vec{k}} \Phi^{s,t}_{\vec{Q},\vec{k}}
    {\ket{\vec{k}}}_{\mathrm{rel}}
    \ket{\chi_{\vec{Q}/2+\vec{k}}^{\vphantom{h}e,s}} \ket{\chi_{\vec{Q}/2-\vec{k}}^{h,{t}}}
     ~, 
\end{equation}
where $\braket{\vec{R}}{\vec{Q}} = e^{i\vec{Q}\boldsymbol{\cdot}{\vec{R}}} / \sqrt{V}$.
The four combinations of the spin-orbit-parity labels $s$ and $t$ lead to four distinct exciton basis states, similarly to the singlet and triplet excitons in regular semiconductors. As explained in the next section, the states $\ket{\vec{Q};s,t}$ in Eq.~\eqref{eq:exc_state_Q} are exact eigenstates of the Wannier problem with only the direct interaction included. This is in fact a much used approximation in the literature \cite{chen2017a}, but in our case not sufficiently accurate as the exchange interaction in principle couples these states for $\vec{Q} \neq \vec{0}$. Nevertheless, the above set of states can still be considered as the most appropriate basis for the full excitonic problem. Further note that, as advanced, $\ket{\vec{Q};s,t}$ more precisely stands for an entire family of ro-vibrational states which must be labeled by additional quantum numbers describing the different relative wave functions that solve the exciton problem.

The topology of these exciton basis states can now be intuitively understood, as in the single-electron case, from the dependence of the expectation value of the pseudospin $\vec{\Gamma}_{e}(\vec{Q})$ on the momentum $\vec{Q}$, obtained from Eq.\,(\ref{eq:exc_state_Q}) as
\begin{equation} \label{eq:gamma_expect}
    \vec{\Gamma}_{e}(\vec{Q}) = \frac{1}{V} \sum_{\vec{k}}
    |\Phi^{s,t}_{\vec{Q},\vec{k}}|^2
    \bra{\chi_{\vec{Q}/2+\vec{k}}^{e,s}} \vec{\Gamma}
    \ket{\chi_{\vec{Q}/2+\vec{k}}^{e,s}} ~,
\end{equation}
\noindent and similarly for the pseudospin of the hole.

To determine the topology of the states $\ket{\vec{Q};s,t}$ mathematically more rigorously, we need to compute the winding number of the pseudospin.
Note that for the moment we particularize to the case of a globally vanishing exchange interaction, in which case all states $\ket{\vec{Q};s,t}$ are uncoupled and therefore have a well-defined Chern number.
To access the topological properties we must compute the Berry connection \cite{shen2012,ortmann2015,asboth2016short,tkachov2015topological,luo2019}, which in the excitonic case is given by a sum of three distinct terms, $\vec{A}_{s,t}(\vec{Q}) = \vec{A}^{(e)}_{s,t}(\vec{Q}) + \vec{A}^{(h)}_{s,t}(\vec{Q}) + \vec{A}^{(\Phi)}_{s,t}(\vec{Q})$.
These read
\begin{align}
    \vec{A}^{(e)}_{s,t}(\vec{Q}) &= -\frac{i}{V} \sum_{\vec{k}}
    |\Phi^{s,t}_{\vec{Q},\vec{k}}|^2
    \bra{\chi_{\vec{Q}/2+\vec{k}}^{e,s}} \vec{\nabla}_{\vec{Q}}
    \ket{\chi_{\vec{Q}/2+\vec{k}}^{e,s}} ~, \\
    \vec{A}^{(h)}_{s,t}(\vec{Q}) &= -\frac{i}{V} \sum_{\vec{k}}
    |\Phi^{s,t}_{\vec{Q},\vec{k}}|^2
    \bra{\chi_{\vec{Q}/2-\vec{k}}^{h,t}} \vec{\nabla}_{\vec{Q}}
    \ket{\chi_{\vec{Q}/2-\vec{k}}^{h,t}} ~, \\
    \vec{A}^{(\Phi)}_{s,t}(\vec{Q}) &= -\frac{i}{V} \sum_{\vec{k}}
    (\Phi^{s,t}_{\vec{Q},\vec{k}})^{*} \,
    \vec{\nabla}_{\vec{Q}}
    \Phi^{s,t}_{\vec{Q},\vec{k}} ~.
\end{align}
Each of these terms contributes to the local Berry curvature or momentum-space magnetic field through $\vec{B}^{(i)}_{s,t}(\vec{Q}) = \vec{\nabla}_{\vec{Q}} \times \vec{A}^{(i)}_{s,t}(\vec{Q})$, where $i = e, h, \Phi$.
Finally, the Berry curvature is connected to the desired winding or Chern number by
\begin{equation} \label{eq:chern_eh}
    \mathcal{C} = \frac{1}{2\pi} \int \mathrm{d}^{2} \vec{Q} \vec{\cdot} \vec{B}(\vec{Q}) = \frac{1}{2 \pi} \oint \mathrm{d} \vec{Q} \vec{\cdot} \vec{A}(\vec{Q}) ~,
\end{equation}
where the first surface integral is performed over the first Brillouin zone, which in our continuum model amounts to integration over the infinite momentum space, and the equivalent line integral is performed along a contour with $Q \rightarrow \infty$.
It is clear that the terms $\vec{B}^{(e)}_{s,t}(\vec{Q})$ and $\vec{B}^{(h)}_{s,t}(\vec{Q})$ not only contain the direct weighted Berry curvature of the underlying single particles, but also an interference term between the single-particle Berry connection and the excitonic envelope wave function.
However, the total contribution of these terms to the exciton Chern number is given only by the single-particle electron or hole Chern number.
This can be shown, e.g., for the electron case, by first expanding $\bra{\chi^{e,s}_{\vec{Q}/2 + \vec{k}}} \vec{\nabla}_{\vec{Q}} \ket{\chi^{e,s}_{\vec{Q}/2 + \vec{k}}}$ with respect to $\vec{k}$ and noticing that $\vec{\nabla}_{\vec{k}} \ket{\chi^{e,s}_{\vec{Q}/2 + \vec{k}}} \big|_{\vec{k} = \vec{0}} = 2 \vec{\nabla}_{\vec{Q}} \ket{\chi^{e,s}_{\vec{Q}/2}}$.
In our model one then observes that all terms in the line integral decay faster with $Q$ than the zeroth-order term $\bra{\chi^{e,s}_{\vec{Q}/2}} \vec{\nabla}_{\vec{Q}} \ket{\chi^{e,s}_{\vec{Q}/2}}$ and thus vanish on the contour at infinity.
Therefore,
\begin{equation}
    \mathcal{C}_{e} = {-} \frac{i}{2 \pi} \oint \mathrm{d} \vec{Q} \vec{\cdot} \bigg(\frac{1}{V} \sum_{\vec{k}} |\Phi^{s,t}_{\vec{Q,\vec{k}}}|^{2} \bra{\chi^{e,s}_{\vec{Q}/2}} \vec{\nabla}_{\vec{Q}} \ket{\chi^{e,s}_{\vec{Q}/2}} \bigg) ~,
\end{equation}
and the normalization condition $\frac{1}{V} \sum_{\vec{k}} |\Phi^{s,t}_{\vec{Q,\vec{k}}}|^{2} = 1$ may now be used.
We are then left precisely with the expression for the free electron Chern number.

There can in principle be an additional contribution to the Chern number due to the envelope wave function itself, given by $\mathcal{C}_{\Phi} = \frac{1}{2 \pi} \oint \mathrm{d} \vec{Q} \vec{\cdot} \vec{A}^{(\Phi)}_{s,t}(\vec{Q})$.
However, this can be shown to vanish here by analyzing the winding of the direct interaction around the origin of $\vec{Q}$.
This winding is trivial, meaning that the potential does not pick up a phase as one circles around this point.
That is, $V^{\mathrm{D}}(\vec{Q}; k, \phi_{\vec{k}}, k', \phi_{\vec{k}'}) = {V^{\mathrm{D}}(Q \hat{\vec{x}}; k, \phi_{\vec{k}} - \phi_{\vec{Q}}, k', \phi_{\vec{k}'} - \phi_{\vec{Q}})}$, where $\phi_{\vec{q}}$ is the angle between the momentum $\vec{q}$ and the $x$-axis.
A straightforward analysis of the exciton eigenproblem then shows that we can choose a gauge such that $\Phi^{s,t}_{\vec{Q}}(k, \phi_{\vec{k}}) = \Phi^{s,t}_{Q \hat{\vec{x}}}(k, \phi_{\vec{k}} - \phi_{\vec{Q}})$.
Observing that the system enjoys reflection symmetry with respect to the $x$-axis when $\vec{Q} = Q \hat{\vec{x}}$, this may now be used to verify that $\mathcal{C}_{\Phi} = 0$.

We thus conclude that all global topological properties of the excitons are introduced by the electron and hole states $\ket{\chi_{\vec{Q}/2+\vec{k}}^{\vphantom{h}e,s}}$ and $\ket{\chi_{\vec{Q}/2-\vec{k}}^{h,{t}}}$.
Hence, the total Chern number of the above exciton basis states is $\mathcal{C} = \mathcal{C}_{e} + \mathcal{C}_{h}$, which for our BHZ model becomes $\mathcal{C} = s + t$ by explicit calculation.
Physically, this result can be most easily understood by the fact that the even and odd spin-orbit-parity subspaces are related by time-reversal symmetry and that the wave function for the hole is the complex conjugate of the electronic valence-band wave function, as we have seen.
Note, however, that the local properties that influence for instance the electron and hole transport are quantified by the Berry curvature, and are thus still dependent on the precise shape of the relative wave functions and the interference terms.

We stress that the intuitive picture given above is only valid in the case of zero exchange interaction.
As seen in the following section, when this is included the two subspaces with $s = t$ become coupled together and cannot be treated individually.
The Chern number is then technically not well-defined since the time-reversal symmetry protects the degeneracy at $\vec{Q} = \vec{0}$.
However, the nontrivial winding caused by the underlying Chern numbers remains, and as a result the full excitonic eigenstates still possess a chirality of $\pm 2$.
This is crucially dependent on the fact that the exchange interaction that couples the two subspaces winds nontrivially around the origin of $\vec{Q}$, transforming instead as ${V^{\mathrm{X}}(\vec{Q}; k, \phi_{\vec{k}}, k', \phi_{\vec{k}'}) = e^{{-}2 i \phi_{\vec{Q}}} V^{\mathrm{X}}(Q \hat{\vec{x}}; k, \phi_{\vec{k}} - \phi_{\vec{Q}}, k', \phi_{\vec{k}'} - \phi_{\vec{Q}})}$.

\subsection{Electron-hole interaction and Bethe-Salpeter equation}\label{sec:4}

Having introduced the free part of the full Hamiltonian, we now proceed to discuss the electron-hole interaction potential that binds these particles together to form an exciton state.
The interaction potential is $V_{s,t;s',t'}(\vec{Q};\vec{k},\vec{k}') = V^{\mathrm{D}}_{s,t;s',t'}(\vec{Q};\vec{k},\vec{k}') - V^{\mathrm{X}}_{s,t;s',t'}(\vec{Q};\vec{k},\vec{k}')$, where $V^\mathrm{D}$ and $V^\mathrm{X}$ denote the direct and exchange interactions, respectively.
These are given by
\begin{align}
    \label{eq:eff_cou_mat_elem_A}
        V^{\mathrm{D}}_{s,t;s',t'}&(\vec{Q};\vec{k},\vec{k}') = \delta_{s,s'} \delta_{t,t'} V(\vec{k} - \vec{k'}) \braket{\chi_{\vec{Q}/2+\vec{k}}^{c,{s}}}{\chi_{\vec{Q}/2+\vec{k}'}^{c,{s}}} \braket{\chi_{-\vec{Q}/2+\vec{k}'}^{v,{t}}}{\chi_{-\vec{Q}/2+\vec{k}}^{v,{t}}}
     ~, \\
    \label{eq:eff_cou_mat_elem_B}
        V^{\mathrm{X}}_{s,t;s',t'}&(\vec{Q};\vec{k},\vec{k}') = \delta_{s,t} \delta_{s',t'} V(\vec{Q}) \braket{\chi_{\vec{Q}/2+\vec{k}}^{c,{s}}}{\chi_{-\vec{Q}/2+\vec{k}}^{v,s}} \braket{\chi_{-\vec{Q}/2+\vec{k}'}^{v,{s'}}}{\chi_{\vec{Q}/2+\vec{k}'}^{c,s'}}
     ~,
\end{align}
where $V(\vec{q})$ is the bare electrostatic potential within the nanosheet. A variational approach with the trial wave functions in the previous section leads to the Bethe-Salpeter equation \cite{martin2016,stoof2009}
\begin{equation}
  \sum_{\boldsymbol{k}', s', t'} \hspace{1.2mm} \langle \boldsymbol{Q}, \boldsymbol{k}; s, t | H | \boldsymbol{Q}, \boldsymbol{k}'; s', t'\rangle \hspace{0.3mm} \Phi_{\boldsymbol{Q}, \boldsymbol{k}'}^{s',t'} = \varepsilon^{\vphantom{s,t}}_{\boldsymbol{Q}} \Phi_{\boldsymbol{Q}, \boldsymbol{k}}^{s,t} ~,
\end{equation}
where we have restored the spin-orbit-parity indices in the wave functions. The matrix elements of the total Hamiltonian, including the electron-hole interaction, read
\begin{equation}\label{eq:matrixEl}
    \langle \boldsymbol{Q}, \boldsymbol{k}; s, t | H | \boldsymbol{Q}, \boldsymbol{k}'; s', t'\rangle = \big[\epsilon_{c}(\boldsymbol{Q}/2 + \boldsymbol{k}) - \epsilon_{v}(\boldsymbol{Q}/2 - \boldsymbol{k})\big] \delta_{\boldsymbol{k},\boldsymbol{k}'} \delta_{s,s'} \delta_{t,t'} + \frac{1}{V} \big[ V^{\mathrm{D}} _{s,t;s',t'}(\boldsymbol{Q}; \boldsymbol{k}, \boldsymbol{k}') - V^{\mathrm{X}}_{s,t;s',t'}(\boldsymbol{Q}; \boldsymbol{k}, \boldsymbol{k}') \big] ~.
\end{equation}

We have solved this equation with both the bare two-dimensional Coulomb potential $V^{\mathrm{C}}(\vec{q})$ with the dielectric constant of the surrounding medium $\varepsilon_{s}$, and with the \HN potential $V^{\mathrm{RK}}(\vec{q})$, which also takes into account the dielectric constant $\varepsilon_{d}$ of Bi\textsubscript{2}Se\textsubscript{3} and is typically used in a nanosheet geometry. The explicit forms of these potentials in momentum space read
\begin{align}
    V^{\mathrm{C}}(\vec{q}) &= {-}\frac{e^{2}}{2 \varepsilon_{0} \varepsilon_{s} q} ~, \\
    V^{\mathrm{RK}}(\vec{q}) &= {-}\frac{e^{2}}{2 \varepsilon_{0} \varepsilon_{s}} \frac{1}{q(1 + r_{0}q)} ~,
\end{align}
where $r_{0} = (\varepsilon_{d}/2 \varepsilon_{s}) L_{z}$ is a screening length that depends on the dielectric constants of Bi\textsubscript{2}Se\textsubscript{3} and the surrounding environment.
We have chosen the relative permittivity ${\varepsilon_{s} = 6}$, which is a typical low value for the environment \cite{feng2019dielectric}.
The bulk dielectric constant has been set to ${\varepsilon_{d} = 28}$ in accord with recent first-principles studies of rhombohedral Bi\textsubscript{2}Se\textsubscript{3} in the near-infrared region \textcolor{black}{at 6 QLs} \cite{cao2013topological,fang2020layer}.

Our approach implies that we neglect the effects of quantum confinement on the classical electrostatic interaction.
\textcolor{black}{This is acceptable if the in-plane separation of the bound electron and hole is larger than the nanosheet thickness, as in this case the electric field lines will mostly lie in the surrounding environment.}
We find that the long-wavelength Coulomb interaction alone cannot accurately model excitons in bismuth selenide nanosheets because the resulting exciton diameters of the low-lying states are in fact smaller than the slab thickness, as seen in Table~\ref{tbl:chExcBi2Se3:meanExcitonDiameter}.
Coincidentally, the \HN interaction closely resembles the quantum-confined Keldysh potential at all momenta \cite{garciaflorez2020}, which is in any case guaranteed to accurately model the electrostatic interaction in this geometry.
For this reason, it is not important which of these potentials enters Eqs.~\eqref{eq:eff_cou_mat_elem_A} and \eqref{eq:eff_cou_mat_elem_B}, and we have additionally checked that the difference in binding energies obtained with the \HN and the quantum-confined Keldysh potentials are no larger than $\SI{1}{\milli\electronvolt}$, leading to errors lower than $10\%$.
Note, however, that fully incorporating the quantum confinement would in principle lead to electrostatic interactions between subspaces with different spin-orbit parity due to the small overlap of the wave functions in the $z$-direction.
This may have a small effect on the \HN ground-state excitons, which are slightly smaller than the nanosheet thickness, but we note that the dielectric constant we have employed for the surrounding environment is relatively low and using a higher value would also lead to bigger excitons and thus mitigate this issue.

Additionally, we must compare the binding energies of the excitons with the splitting of the first two bulk subbands due to confinement, as we have neglected the subspaces of higher energy.
\textcolor{black}{Our approximation of projecting the 3D Hamiltonian onto the bulk subbands closest to the Fermi surface will be justified if the former is smaller than the latter.
A binding energy larger than the bulk-subband splitting would imply the need to include the wave functions of higher excited states and thus effectively hinder the two-dimensional treatment of the problem.}
The splittings of the conduction and valence subbands for $L_{z} = \SI{6}{\nano\meter}$ are found to be about $\SI{61}{\milli\electronvolt}$ and $\SI{24}{\milli\electronvolt}$, respectively, and the binding energies are given in Fig.~\ref{fig:exc_wave function}.
Except for the ground state, all states have a binding energy that is smaller than the relevant subband splittings.
In the case of the ground state, the binding energy is only around $\SI{1}{\milli\electronvolt}$ larger than the valence-band splitting, so we do not expect that including the second subband will lead to significant modifications.
We conclude that the \HN potential is apt for our study of excitons in Bi\textsubscript{2}Se\textsubscript{3} nanosheets, especially if we use a higher relative permittivity for the environment.
Hence, unless otherwise specified, all figures in this article correspond to results obtained via the \HN potential.
Note, however, that the topology is neither affected by the values of the dielectric constants, nor by the precise form of the interaction potential.

\begin{table}[!htb]
   \centering
   \begin{tabular}{ccccccccc}
   \multicolumn{2}{c|}{$m$}                 & $0$ & $-1$ & $+1$ & $-2$ & $+2$ & $-3$ & $+3$ \\
   \hline
   \multicolumn{1}{l|}{} & \multicolumn{1}{l|}{$V^{\mathrm{C}}$} & $2.74$ & $4.21$ & $5.32$ & $6.33$ & $7.61$ & $9.04$ & $9.76$      \\ \cline{2-9}
   \multicolumn{1}{l|}{\multirow{-2}{*}{$\sqrt{\langle r^{2}\rangle}$ (\si{\nano\meter})}}                        & \multicolumn{1}{l|}{$V^{\mathrm{RK}}$} & $5.64$ & $7.02$ & $7.51$ & $8.72$ & $9.13$ & $10.57$ & $10.89$ \\  
   \end{tabular}
   \caption{Mean exciton diameters of the zero-momentum states $\ket{\vec{0};+,+;0,m}$ and $\ket{\vec{0};-,-;0,-m}$ as obtained with the Coulomb and \HN potentials with $\varepsilon_{s} = 6$ and $\varepsilon_{d} = 28$. In the case of the Coulomb potential the radii must be compared to the film thickness, which is $L_{z} = \SI{6}{\nano\meter}$.}
   \label{tbl:chExcBi2Se3:meanExcitonDiameter}
\end{table}

Due to the orthogonality of the states in the different spin-orbit-parity subspaces the exchange interaction $V^{\mathrm{X}}$ only contributes when the spin-orbit parities of the electron and the hole are equal in both the initial and final states and when $\vec{Q}\neq\vec{0}$, as the inner products in Eq.\,(\ref{eq:eff_cou_mat_elem_B}) tend linearly to zero as a function of $\vec{Q}$.
Thus, specifically, the states $\ket{\vec{Q};+,-}$ and $\ket{\vec{Q};-,+}$ are not coupled by the exchange interaction, whereas the states $\ket{\vec{Q};+,+}$ and $\ket{\vec{Q};-,-}$ are. We can therefore solve the problem in the subspaces spanned by $\ket{\vec{Q};+,+}$ and $\ket{\vec{Q};-,-}$ on the one hand, and by $\ket{\vec{Q};\pm,\mp}$ on the other hand. Since the interaction potentials are, up to complex conjugation, the same for the states $\ket{\vec{Q};+,-}$ and $\ket{\vec{Q};-,+}$, the corresponding wave functions also only differ by complex conjugation, and both states are degenerate in energy for all $\vec{Q}$. By contrast, the eigenstates spanned by $\ket{\vec{Q};+,+}$ and $\ket{\vec{Q};-,-}$ are degenerate in energy only for $\vec{Q} = \vec{0}$. 


\subsection{Derivation of optical properties}

The oscillator strength of an exciton $\ket{\vec{0}; s, t; n, m}$ reads \cite{zhang2018optical,cao2018}
\begin{equation}
    f^{m, \hat{\vec{e}}}_{s,t;n} = \frac{2}{\varepsilon^{m}_{s,t;n}} \bigg|\int \mathrm{d}^{2} k \, \Phi^{(m)}_{\vec{0}; s,t;n} \, e^{i m \phi_{\vec{k}}} \, \hat{\vec{e}} \vec{\cdot} \bra{\chi^{v,t}_{\vec{k}}} \vec{v}(\vec{k}) \ket{\chi^{c,s}_{\vec{k}}}\bigg|^{2} ,
\end{equation}
where $\hat{\vec{e}}$ is the Jones vector of the outgoing beam, $\varepsilon^{m}_{s,t;n}$ is the energy of the zero-momentum exciton, and $\vec{v}(\vec{k})$ is the velocity operator. Since only interatomic transitions are expected to play a role in Bi\textsubscript{2}Se\textsubscript{3}, the velocity operator is well approximated by $\vec{v}(\vec{k}) \approx \vec{\nabla}_{\!\vec{k}} H_{0}(\vec{k})$ \cite{pedersen2001optical}. The Jones vectors of left- and right-circularly polarized light are $\hat{\vec{e}}_{+}$ and $\hat{\vec{e}}_{-}$, respectively, with $\hat{\vec{e}}_{\pm} = \frac{1}{\sqrt{2}}(\hat{\vec{x}} \pm i \hat{\vec{y}})$.
Computing the matrix elements that enter the above equation readily yields the reported results.

\subsection{Effective exciton Hamiltonian}

The small-$\boldsymbol{Q}$ behavior of excitons can be understood by constructing an effective model for the pairs of states $\ket{\boldsymbol{Q};+,+;n, m}$ and $\ket{\boldsymbol{Q};-,-;n, -m}$ that are degenerate at zero momentum, as already done before~\cite{yu2014a, wu2015a,qiu2015nonanalyticity}.
By computing the relevant matrix elements with the wave functions for $\vec{Q} \rightarrow \vec{0}$ we obtain the effective Hamiltonian
\begin{equation}\label{eq:effectiveH}
   H_{X}^{\text{eff}}(\boldsymbol{Q}) = \bigg(\hbar \omega_{X} + \frac{\hbar^2 Q^{2}}{2M_{X}}\bigg)  + \mathcal{J}_{X}(\boldsymbol{Q}) \begin{bmatrix}
      1 & e^{-2i\phi_{\boldsymbol{Q}}} \\ e^{2i\phi_{\boldsymbol{Q}}} & 1
   \end{bmatrix} ~,
\end{equation}
where $X = (n,m)$ labels the particular exciton doublet.
Here, $\hbar \omega_{X}$ is the energy of the doublet at zero momentum, $M_{X}$ is an effective mass, and
\begin{equation}\label{eq:effectiveJ}
       \mathcal{J}_{X}(\boldsymbol{Q}) = \begin{dcases}
      0 & \text{if $m$ is odd} ~, \\
      J_{X} Q + \frac{\hbar^{2} Q^{2}}{2 M'_{X}} + \mathcal{O}(Q^{3}) & \text{if $m$ is even} ~.
   \end{dcases}
\end{equation}
We note that these parameters depend on which particular doublet the model attempts to describe.
According to Eq.~\eqref{eq:effectiveJ}, the states with odd angular momentum at $\vec{Q} = \vec{0}$ should remain degenerate for small $\vec{Q}$, which is indeed the case in Figs.~\ref{fig:exc_disp} and \ref{fig:exc_disp_C}.
However, this degeneracy is broken to third order in $Q$, but this effect is not captured by our perturbative scheme and requires including corrections to the zero-momentum wave functions as well.
On the other hand, the states with even angular momentum may split into a linear mode and a quadratic mode, as is the case for the lowest $s$-wave and $d$-wave states in the figures.
This effect is expected to be most important for states with $m = 0$, as the lowest-order contribution to the splitting parameter $J_{X}$ is proportional to $|\Phi_{\vec{Q}}(\vec{r} = \vec{0})|^{2}$, which is nonzero only for $s$-wave excitons.

To solve this model we restrict ourselves to $s$-wave excitons and neglect the terms proportional to the identity matrix, which do not contribute to the topological properties. Also, we drop the remaining terms proportional to $Q^{2}$, as we are interested in the low-energy physics.
We introduce a small Zeeman-like term $\Delta \sigma_{z}$ that breaks the time-reversal symmetry and makes the model gapped, resulting in $H^{\mathrm{eff}}(\vec{Q}) = \vec{b}(\vec{Q}) \vec{\cdot} \vec{\sigma}$ with $\vec{b}(\vec{Q}) = (J Q \cos 2 \phi_{\vec{Q}}, \, J Q \sin 2 \phi_{\vec{Q}}, \, \Delta)$.
The winding number of the Hamiltonian is computed via the formula
\begin{equation}
    w = \frac{1}{4 \pi} \int \mathrm{d}^{2} Q \,\, \hat{\vec{b}} \vec{\cdot} \bigg(\frac{\partial \hat{\vec{b}}}{\partial Q_{x}} \times \frac{\partial \hat{\vec{b}}}{\partial Q_{y}} \bigg) ~,
\end{equation}
and gives $w = \operatorname{sgn} \Delta$.
The $\Delta$ term also introduces Berry curvatures $\Omega^{\pm}_{xy}(\vec{Q}) = \pm J^{2} \Delta / \big(\Delta^{2} + J^{2} Q^{2}\big)^{3/2}$ for the positive- and negative-energy solutions, which give Chern numbers $\mathcal{C} = \pm \operatorname{sgn} \Delta$.

To solve the model with a position-dependent Zeeman term we must perform the substitution $Q_{y} \rightarrow {-}i\partial_{y}$ in the effective Hamiltonian due to the breaking of translational invariance in the $y$-direction.
This introduces some complications, as the object $Q e^{2 i \phi_{\vec{Q}}}$ does not have a unique operator generalization.
However, for ${Q_{x} = 0}$ we unambiguously have $e^{2 i \phi_{\vec{Q}}} = -1$, whereas $Q$ reduces to $\sqrt{\smash[b]{-\partial_{y}^{2}}}$.
We interpret this as an operator whose square is $-\partial_{y}^{2}$ and whose eigenfunctions are plane waves.
However, there are still two possibilities for the action of this operator on a plane wave, namely $\sqrt{\smash[b]{-\partial_{y}^{2}}} e^{\Lambda y} = \pm i \Lambda e^{\Lambda y}$.
These two choices precisely give the two modes we are looking for.
For $Q_{x} = 0$, the model to solve is thus
\begin{equation}\label{eq:effEdgeStates}
    \begin{bmatrix}
        \Delta(y) & {-} J \sqrt{\smash[b]{-\partial_{y}^{2}}} \\ {-} J \sqrt{\smash[b]{-\partial_{y}^{2}}} & -\Delta(y)
    \end{bmatrix} \begin{bmatrix}
        \Psi_{1}(y) \\ \Psi_{2}(y)
    \end{bmatrix} = E \begin{bmatrix}
        \Psi_{1}(y) \\ \Psi_{2}(y)
    \end{bmatrix} ,
\end{equation}
and we look for solutions with $E = 0$ consistent with the sought-after chiral excitons.
For each choice of operator action, the ansatz $\Psi_{1}(y) = e^{\Lambda y}$ and $\Psi_{2}(y) = \chi \Psi_{1}(y)$, with $\chi \in \mathbb{C}$, gives a single solution of the Jackiw-Rebbi type \cite{jackiw1976solitons,shen2012}.
In this way we find the two linearly independent eigenfunctions of Eq.~\eqref{eq:WFedge}.
For small but nonzero $Q_{x}$ we can use perturbation theory to find the first-order correction to the spectrum by using $Q e^{2 i \phi_{\vec{Q}}} = -\sqrt{\smash[b]{Q_{y}^{2}}} \big[1 - 2 i \frac{Q_{x}}{Q_{y}} + \mathcal{O}(Q_{x}^{2})\big]$, where in analogy to the above we define $\frac{1}{\partial_{y}} e^{\Lambda y} = \frac{1}{\Lambda} e^{\Lambda y}$ as the action of the inverse derivative operator on plane-wave solutions.

This analytic solution has been found by neglecting the terms quadratic in $\vec{Q}$ in the effective exciton Hamiltonian.
However, we expect the presence of exciton edge states even when including these higher-order terms, although an explicit calculation is much more involved.
Firstly, we note that the inclusion of the Zeeman-like term will produces a globally gapped exciton spectrum.
This is true not only for the effective model with or without quadratic corrections, but also for the full excitonic spectra of Figs.~\ref{fig:exc_disp} and \ref{fig:exc_disp_C}.
The presence of edge states is then inferred directly from a nonzero Chern number.
Including the quadratic terms yields the following Berry curvatures for the valence and conduction bands:
\begin{align}
    \Omega_{xy}^{\pm}(\vec{Q}) = \pm 4 \Delta |M'| \frac{Q (JM'+Q) (2JM'+Q)}{\big[Q^{2}(2JM'+Q)^{2}+(2M' \Delta)^{2}\big]^{3/2}} ~ .
\end{align}
Integration of this expression yields $\mathcal{C}_{\pm} = \pm \operatorname{sgn} \Delta$, the same as before, so edge modes are expected in this full model as well.

As an additional check, we further note that in this more general case the Berry curvature is nonzero even for $J = 0$.
Hence the topology remains even in this case, for which the Hamiltonian is parabolic and only contains integer powers of $Q_{x}$ and $Q_{y}$.
Thus, we may use the argument of Ref.\cite{mong2011edge} to infer the existence of topological edge states.
The Hamiltonian can be written as $H(\vec{Q}) = b_{0}(\vec{Q}) + \vec{b}(\vec{Q}) \vec{\cdot} \vec{\sigma}$, where for each fixed $Q_{x}$ the vector $\vec{b}(\vec{Q})$ is a parabolic function of the perpendicular momentum $Q_{y}$.
For all momenta $Q_{y} \neq 0$ the origin lies on the concave side of the projection of this parabola onto the plane, and so there should be edge states, which will be chiral due to the nonvanishing topological invariant.

\subsection{Computational details}

The Bethe-Salpeter equation has been solved with an independently developed MATLAB code that implements the discretization procedure of Ref.~\cite{karr2010} after rewriting all equations in terms of the dimensionless momentum $u = k L_{z}$ and keeping only contributions from angular momenta $|m| \leq 3$.
The integrals in the matrix elements have been performed up to a momentum cutoff $U = 10$ and with a discretization $\Delta u = 0.05$, and we have verified that the results do not depend on the cutoff by performing additional calculations up to $U = 40$.
The long-wavelength divergence of the Coulomb potential has been regularized via an infrared cutoff $\Delta_{V} u$ chosen such that $\Delta_{V} u / \Delta u \approx 0.2262$, for which we have checked that the energies do not depend on $\Delta u$.
If $\Delta_{V} u / \Delta u$ is chosen differently, a well-defined extrapolation of the exciton energy levels for $\Delta u \rightarrow 0$ can be done by using different values of the discretization.

In order to identify the angular momentum quantum numbers unambiguously after solving the Bethe-Salpeter equation, we choose a nonsingular gauge for the single-particle eigenstates that enter the direct and exchange potentials \cite{zhou2015berry,zhang2018optical}. We have checked that with this choice the level ordering in the trivial regime reduces to that of the 2D hydrogen atom.


%
%

%


\bibliographystyle{ieeetr}
\bibliography{bibliography}

\section*{Acknowledgments}

We thank Daniel Vanmaekelbergh for many stimulating discussions. This work is part of the research programme TOP-ECHO with project number 715.016.002, and is also supported by the D-ITP consortium. Both are programmes of the Netherlands Organisation for Scientific Research (NWO) that is funded by the Dutch Ministry of Education, Culture and Science (OCW).

\section*{Author contributions}

The work was conceptualized by all four authors.
The analytical and numerical computations were performed by L.M.L., with input of H.S. and F.G.F.
The suggestions for future experiments were provided by L.S.
All authors contributed to the writing of the manuscript.

\section*{Data availability}

The data that supports the findings of this study is available from the corresponding author upon reasonable request.

\section*{Additional information}

\subsection*{Competing interests}

The authors declare no competing interests.

\subsection*{Code availability}

The code that was employed to obtain the numerical results of this study is available from the corresponding author upon reasonable request.

\subsection*{Materials and correspondence}

General correspondence may be sent to L.M.L. or H.S. For data and code requests, contact L.M.L.

\end{document}